\documentclass[structabstract]{aa}  

\usepackage{graphicx}
\usepackage{txfonts}
\usepackage{natbib}
\bibpunct{(}{)}{;}{a}{}{,}
\newcommand{\lsi}{\object{LS~I~+61~303}}
\newcommand{\psr}{\object{PSR~B1259--63}}
\newcommand{\ls}{\object{LS~5039}}
\newcommand{\hess}{\object{HESS~J0632+057}}
\begin{document}

\title{Search for radio pulsations in \object{LS~I~+61~303}}

\author{A.~Ca\~nellas\inst{1}
          \and
          B.~C.~Joshi\inst{2}
          \and
          J.~M.~Paredes\inst{1}
          \and
          C.~H.~Ishwara-Chandra\inst{2}
          \and
          J.~Mold\'on\inst{1}
          \and
          V.~Zabalza\inst{1,3}
          \and
          J.~Mart\'{\i}\inst{4}
          \and
          M.~Rib\'o\inst{1}
          }

\institute{Departament d'Astronomia i Meteorologia, Institut de Ci\`encies del Cosmos (ICC), Universitat de Barcelona (IEEC-UB), Mart\'{\i} i Franqu\`es 1, E08028 Barcelona, Spain\\
           \email{acanyelles@am.ub.es; jmparedes@ub.edu; jmoldon@am.ub.es; vzabalza@am.ub.es; mribo@am.ub.es}
           \and
           National Centre for Radio Astrophysics, Post Bag 3, Ganeshkhind, Pune 411 007, India\\
           \email{bcj@ncra.tifr.res.in; ishwar@ncra.tifr.res.in}
           \and
           Max-Planck-Institut f\"ur Kernphysik, Saupfercheckweg 1, 69117, Heidelberg, Germany\\
           \email{Victor.Zabalza@mpi-hd.mpg}
           \and
           Departamento de F\'{\i}sica (EPSJ), Universidad de Ja\'en, Campus Las Lagunillas s/n, 23071 Ja\'en, Spain\\
           \email{jmarti@ujaen.es}
           }

\abstract
   {\lsi\ is a member of the select group of gamma-ray binaries: galactic binary systems that contain a massive star and a compact object, show a changing milliarcsecond morphology and a similar broad spectral energy distribution (SED) that peaks at MeV-TeV energies and is modulated by the orbital motion. The nature of the compact object is unclear in \lsi , \ls\ and \hess , whereas \psr\ harbours a 47.74\,ms radio pulsar.}
   {A scenario in which a young pulsar wind interacts with the stellar wind has been proposed to explain the very high energy (VHE, $E$\,$>$\,100\,GeV) gamma-ray emission detected from \lsi , although no pulses have been reported from this system at any wavelength. We aim to find evidence of the pulsar nature of the compact object.}
   {We performed phased array observations with the Giant Metrewave Radio Telescope (GMRT) at 1280\,MHz centred at phase 0.54. Simultaneous data from the multi-bit phased array (PA) back-end with a sampling time of $t_\mathrm{samp}$\,$=$\,$128\,\mathrm{\mu s}$ and from the polarimeter (PMT) back-end with $t_\mathrm{samp}$\,$=$\,$256\,\mathrm{\mu s}$ where taken.}
   {No pulses have been found in the data set, with a minimum detectable mean flux density of $\sim$\,0.38\,mJy at 8-$\sigma$ level for the pulsed emission from a putative pulsar with period $P$\,$>$2\,ms and duty cycle $D$\,$=$10\% in the direction of \lsi .}
   {The detection of posible radio pulsations will require deep and sensitive observations at frequencies $\sim$\,0.5--5\,GHz and orbital phases 0.6$-$0.7. However, it may be unfeasible to detect pulses if the putative pulsar is not beamed at the Earth or if there is a strong absorption within the binary system.}

%\abstract{Abstract}

\keywords{
binaries: close --
gamma rays: stars --
pulsars: general --
radio continuum: stars --
stars: individual: \object{LS~I~+61~303} --
X-rays: binaries
}

\maketitle

%%%%%%%%%%%%%%%%%%%%%%%%%%%%%%%%%%%%%%%%%%%%%%%%%%%%%%%%%%%%%%%%%%%%
\section{Introduction} \label{sec:intro}

Located at a distance of $2.0\pm0.2\,\rm{kpc}$ \citep{frail91}, \lsi\ contains a rapidly rotating B0\,Ve star with a stable equatorial shell, and a compact object of unknown nature with a mass between 1 and 5\,$\mathrm{M}_{\sun}$, orbiting it every 26.5\,days \citep{hutchings81,casares05a}. Optical and IR orbital modulation have been found \citep{mendelson94,paredes94}. Using radial velocities data, \citet{casares05a} found that the orbit is eccentric ($e$\,$\simeq$\,0.72) and periastron takes place at phase $0.23\pm0.02$, taking $T_0$\,$=$\,$\rm{JD}\,2,443,366.775$. More recently \citet{aragona09} stated that the eccentricity is slightly lower ($e$\,$\simeq$\,0.54) and the phase of the periastron passage is 0.275. The compact object would be a neutron star for inclinations $25\degr$\,$\la$\,$i$\,$\la$\,$60\degr$ and a black hole if $i$\,$\la$\,$25\degr$ \citep{casares05a}. However, there are discrepancies between the proposed orbital solutions by \citet{casares05a} and \citet{aragona09}, and the inclination angle is poorly constrained by radial velocities of the companion star alone.

Quasi-periodic radio outbursts monitored during 23 years have provided an accurate orbital period value of $26.4960\pm0.0028\,\rm{d}$ \citep{gregory02}. The maximum of the radio outbursts varies between phase 0.45 and 0.95.

Orbital X-ray periodicity has also been found \citep{paredes97,harrison00,torres10}. Similar results have been obtained at higher energies with \textit{INTEGRAL} data \citep{hermsen06,zhang10}. \lsi\ is also spatially coincident with a high energy (HE, $E$\,$>$\,100\,MeV) gamma-ray source detected by EGRET \citep{kniffen97}.

The \textit{Fermi Space Telescope} Large Area Telescope (\textit{Fermi}/LAT) reported the first detection of the orbital modulation at HE with a period of $26.6\pm0.5$\,days roughly anti-correlated with the X-ray emission \citep{abdo09a}.

At very high energy (VHE, $E$\,$>$\,100\,GeV) gamma rays, \lsi\ was detected by the MAGIC Cherenkov telescope \citep{albert06} and confirmed by the VERITAS stereoscopic array \citep{acciari09b}. Further observations by the MAGIC collaboration led to the discovery of the orbital variability of TeV emission with a period of $26.8\pm0.2$\,\rm{days} \citep{albert09}. Simultaneous observations of MAGIC and the \textit{XMM-Newton} and \textit{Swift} X-ray satellites revealed a correlation between the X-rays and VHE bands \citep{anderhub09}.

\citet{massi04} reported the discovery of an extended jet-like and precessing radio emitting structure at angular extensions of 10--50 milliarcseconds. Owing to the presence of apparently relativistic radio emitting jets, \lsi\ was proposed to be a microquasar. However, VLBA images obtained during a full orbital cycle show a rotating elongated morphology \citep{dhawan06}, which may be consistent with a model based on the interaction between the relativistic wind of a young non-accreting pulsar and the wind of the donor star \citep{maraschi81,dubus06}. On 2008 September 10, the \textit{Swift} Burst Alert Telescope (BAT) detected a hard X-ray burst in the direction of \lsi\ which, assuming the association, would be the signature of a magnetar-like activity that has been proposed to be linked to the presence of a young highly magnetized pulsar in the binary system \citep{dubus08}. The \textit{Fermi}/LAT spectrum is compatible with a power law and an exponential cutoff at $\sim$\,6\,GeV, suggesting that there are two separate spectral components for the HE and VHE emission. Moreover, the spectral similarity with gamma-ray pulsars leads to the consideration of a magnetospheric origin for this HE component despite the fact that no pulses have been detected and although the orbital modulation would be unexpected in this scenario \citep{abdo09a, petri11}. This orbital variability could be understood in the framework of inverse Compton scattering of photon fields in a striped pulsar wind model, which predicts pulsed and variable HE emission \citep{petri11}. In any case, from an observational point of view it is not clear yet if \lsi\ contains an accreting black hole, an accreting neutron star or a non-accreting neutron star.

In addition to \lsi , three binary systems that contain a massive star and a compact object and display extended radio emission have been clearly detected at VHE: \psr , \ls\ and \hess . These four systems have a similar spectral energy distribution (SED; \citealt{dubus06, hinton09}), peaking at MeV-GeV energies, and thus are considered gamma-ray binaries. Whereas the nature of the compact object in \lsi , \ls\ and \hess\ remains unknown, \psr /\object{LS~2883} is the only gamma-ray binary with a confirmed pulsar \citep{johnston92}. This binary system contains an O9.5\,Ve \citep{negueruela11} donor and a 47.74\,ms radio pulsar orbiting it every 3.4\,years in a very eccentric orbit with $e$\,$=$\,$0.87$ \citep{johnston92,johnston94}. The radio light curve of the unpulsed emission of \psr\ is well explained by the adiabatic expansion of a synchrotron bubble, which is similar to the behaviour found in \lsi . Pulsed radio emission is detected with a spectral index of about $-0.6$, although the radio pulses vanish for some weeks during the periastron passage \citep{connors02,johnston05} probably due to free-free absorption by the stellar wind and interaction with the Oe-star disk. Note that in the case of \lsi\ the apastron separation is smaller than the periastron separation for \psr , therefore detecting pulses from \lsi\ is unlikely even if the pulsar exists and is beamed at the Earth. The system was detected by HESS in the 2004 and 2007 periastron passages \citep{aharonian05,aharonian09}. Extended and variable radio emission at milliarcsecond scales, outside the binary system, was observed after the 2007 periastron passage of \psr\ \citep{moldon11b}. This is the first observational evidence that a non-accreting pulsar orbiting a massive star can produce a milliarcsecond radio structure similar to those of \lsi , \ls\ and \hess .

As in the case of \lsi , the nature of the compact object in \ls\ is unknown: it is either a black hole or a neutron star with a mass between 1.5 and 10\,$\mathrm{M}_{\sun}$ in a slightly eccentric ($e$\,$=$\,0.35) 3.9-day orbit around an O6.5\,V((f)) star located at 2.5\,kpc \citep{casares05b,aragona09}. Similarly to \lsi\ and \psr , \ls\ also shows X-rays, HE and VHE emission modulated by the orbit (\citealt{takahashi09,abdo09b,aharonian06}; respectively). The discovery of a bipolar extended milliarcsecond radio emission morphology with VLBA observations prompted a microquasar interpretation \citep{paredes00}, but additional VLBA observations revealed a changing behaviour of the morphology that cannot easily be explained by a microquasar scenario \citep{ribo08}. \citet{rea11b} performed recently a deep search for pulsations from \ls\ with \textit{Chandra}, finding no periodic signals in a frequency range of 0.005--175\,Hz.

\hess\ has recently joined the short list of gamma-ray binaries. \citet{moldon11a} reported the slightly extended and variable radio milliarcsecond structure of \hess\ at 1.6\,GHz with the e-EVN. It is positionally coincident with a B0pe star and the X-ray source XMMU~J063259.3+054801 \citep{hinton09}, which is also variable \citep{acciari09a}. \citet{bongiorno11} have recently confirmed its binary nature by the discovery of a periodicity of 320\,$\pm$\,5\,days in a 0.3--10\,keV light curve obtained with \textit{Swift}. This faint and point-like VHE source was first detected in the HESS Galactic Plane Survey \citep{aharonian07}. \hess\ is clearly variable also at VHE: 2006-2007 and 2008-2009 observations by VERITAS imposed flux upper limits well below the values detected by the HESS Galactic Plane Survey \citep{acciari09a}, and afterwards higher TeV gamma-ray emission was detected by VERITAS \citep{ong11} and MAGIC \citep{mariotti11}. No pulses have been detected at any wavelength from \hess\ \citep{rea11a}.

Overall, there are many similarities between \psr\ and the other three known gamma-ray binaries (\lsi , \ls\ and \hess ), but no pulsations have been detected up to now in the latter. In this paper we report our search for radio pulsations from \lsi . In Sect.~\ref{sec:previous} we present the situation of the searches for pulsations from \lsi\ before our work. In Sect.~\ref{sec:absorption} we describe how we selected the orbital phase of the observation and the frequency in order to minimise the effect of absorption and pulse broadening in the observations described in Sect.~\ref{sec:observations}. We describe the data analysis in Sect.~\ref{sec:analysis}. The results are presented in Sect.~\ref{sec:results}, where we describe the implications for future observations. In Sect.~\ref{sec:discussion} we discuss these results and present our conclusions.

\section{Previous searches of pulsations} \label{sec:previous}

To our knowledge, only two limited searches for evidence of a pulsar in \lsi\ were reported in refereed journals before our observations. \citet{coe82} recorded 10.7-GHz radio data with a time resolution of $\sim$\,15\,s, but no evidence of short-term periodicities was found in the range 28--4200\,s. The only observational hint of pulsed emission has been the discovery by \citet{peracaula97} of a $\sim$\,1.4\,h periodicity in a series of $\sim$\,4\,mJy microflares active during 8\,h in a 5\,GHz VLA radio observation. \citet{mcswain11} recently reported on GBT observations placing deep upper limits between 4.1 and 14.5\,$\mu$Jy at C, S and X bands.

\lsi\ does not appear as a radio pulsar in any of the public catalogues built after performing blind pulsar searches. \lsi\ has a relatively high dispersion measure (DM): the NE2001 electron density model \citep{cordesandlazio02} gives $DM$\,$=$\,$60\pm10$\,$\mathrm{pc\,cm^{-3}}$ in the direction and distance to \lsi\ ($d$\,$=$\,$2.0\pm0.2$\,kpc). As shown in Appendix~\ref{app}, the pulse broadening is also considerable at low frequencies.

The Green Bank and Jodrell Bank observatories have performed pulsar surveys pointing to the \lsi\ region. The Green Bank short-period survey conducted by \cite{stokes85} used a sampling time of 2\,ms. These authors computed a minimum detectable flux density of $\sim$\,100\,mJy for a period of $\sim$\,4\,ms, or $\sim$\,2\,mJy for $P$\,$>$\,100\,ms. More recently, the GBT350 Survey of the Northern Galactic Plane for Radio Pulsars and Transients, using a fast sampling time ($t_\mathrm{samp}$\,$=$\,$81.92$\,$\mu\mathrm{s}$), attained a $\sim$\,2\,mJy sensitivity for normal, slow pulsars (\citealt{hessels08}; this sensitivity should be extendible to 10$-$100\,ms pulsars with moderate DMs, J. W. T. Hessels, private communication).

In the case of Jodrell Bank, \citet{clifton86} claimed a sensitivity of $\sim$\,1\,mJy (considering a $6\sigma$ detection threshold) for a pulsar with a duty cycle of 4\% for the Jodrell Bank B survey.

The surveys of Molonglo, Parkes and Swinburne were made from the southern hemisphere, at latitudes where \lsi\ is not visible. \lsi\ is outside the range of available declinations for Arecibo.

At other wavelenghts, deep \textit{Chandra} observations performed by \citet{rea10} did not exhibit X-ray pulses. Despite showing a HE spectrum similar to those observed in gamma-ray pulsars, \textit{Fermi}/LAT has not reported pulsed gamma-ray emission either. However, the lack of good orbital ephemeris and an \textit{a priori} knowledge on the putative pulsar period hinders the search in the currently available \textit{Fermi}/LAT data.

\section{Minimising absorption} \label{sec:absorption}

\lsi\ is an eccentric binary system in which the proximity of the Be star to the compact object should cause a significant free-free absorption of the putative pulsar radio emission. This is strongly modulated along the orbit through the pronounced eccentricity of the system. In our model we consider the effect of free-free absorption with the stellar wind of the massive star, and we ignore the interaction with the circumstellar disk because it is truncated at 34--37~R$_{\odot}$, much lower than the distance at the considered orbital phase. We consider a smooth and continuous stellar wind as a first-order approximation although a realistic wind is affected by turbulence and may be clumpy. We do not discuss other possible influences of the Tsytovitch-Razin effect and the synchrotron self-absorption (see for instance \citealt{hornby66,white95,dougherty03}). Other effects may be important, as discussed in \cite{thompson94}, such as refraction of the radio beam, pulse smearing, and induced Compton scattering. These
effects are unknown due to the lack of information about the putative pulsar properties and the conditions of the plasma surrounding it.

Because we aimed to observe at the orbital phase with a minimum absorption, we computed the optical depth towards the compact object at different orbital phases and for different frequencies using the Rosseland mean opacity (see \citealt{dubus06}):
%-------------------------------------------------------------------
\begin{eqnarray}
\tau \simeq 0.3
&\times & \left[ \frac{\dot{M_\mathrm{W}}}{10^{-8}\,\mathrm{M_{\sun}\,yr^{-1}}}\right] ^{2}
\times \left[ \frac{v_\mathrm{W}}{2000\,\mathrm{km\,s^{-1}}}\right] ^{-2}
\nonumber \\
& \times & \left[ \frac{T}{10^4\,\mathrm{K}}\right] ^{-3/2}
\times \left[ \frac{\nu}{\mathrm{GHz}}\right] ^{-2}
\times \int_{l_0}^\infty r^{-4}dl\ .
\label{eq:tau}
\end{eqnarray}
%-------------------------------------------------------------------
The assumed parameters for the Be star are a mass-loss rate of ${\dot M_\mathrm{w}}$\,$=$\,$10^{-8}\,\mathrm{M_{\sun}\,yr^{-1}}$ \citep{howarth89}, a terminal wind speed of $v_{\mathrm{w}, \infty}$\,$=$\,$1750\,\mathrm{km\,s^{-1}}$ \citep{hutchings79} and a $\beta$-law with $\beta$\,$=$\,0.8 \citep{puls96}, and a temperature of $T$\,$=$\,28\,000\,K \citep{casares05a}. Distances are in astronomical units. We integrated along the line-of-sight path from the position of the compact object ($l_0$) up to the observer at $\infty$. We have considered the orbital parameters in \citet{casares05a}. Since we are assuming that the compact object is a neutron star, we used 1.4\,$\mathrm{M}_{\sun}$ for its mass. Hence, the mass function implies an inclination angle $i=65\degr$, which is slightly out but not very far of the rough limit of 60\,$\degr$ imposed by the lack of clear shell lines in \citet{casares05a}. Note that these assumptions in the stellar wind and the choice of the orbital parameters plays an important role in the resulting absorption.

\begin{figure}
\includegraphics[]{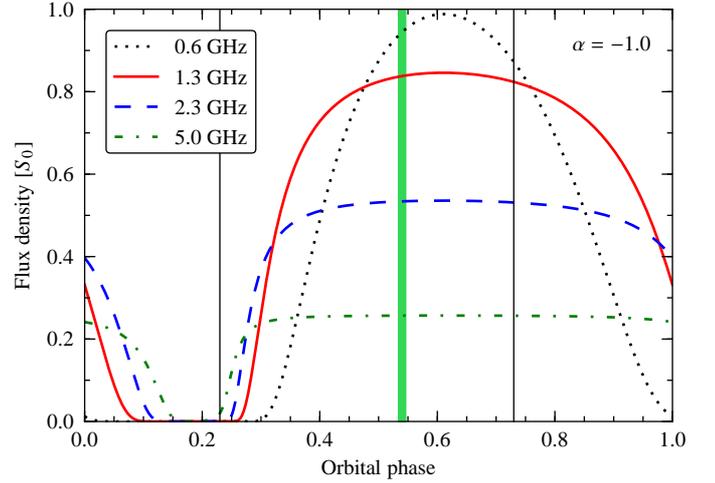}
\caption{Expected \lsi\ pulsed flux density along the orbit at different frequencies for $\alpha$\,$=$\,$-1.0$ assuming that the intrinsic pulsed flux density at 1.3\,GHz is $S_0$ and considering free-free absorption effects. Periastron and apastron are indicated by the vertical black lines at phases 0.23 and 0.73, respectively. The highest observable flux density is expected to occur at 0.6\,GHz at orbital phase $\sim$\,0.6$-$0.7, three days before apastron. The observation phase range of this work is indicated by the vertical green stripe centred at 0.54.}
\label{fig:abs}
\end{figure}

\begin{figure*}
\includegraphics{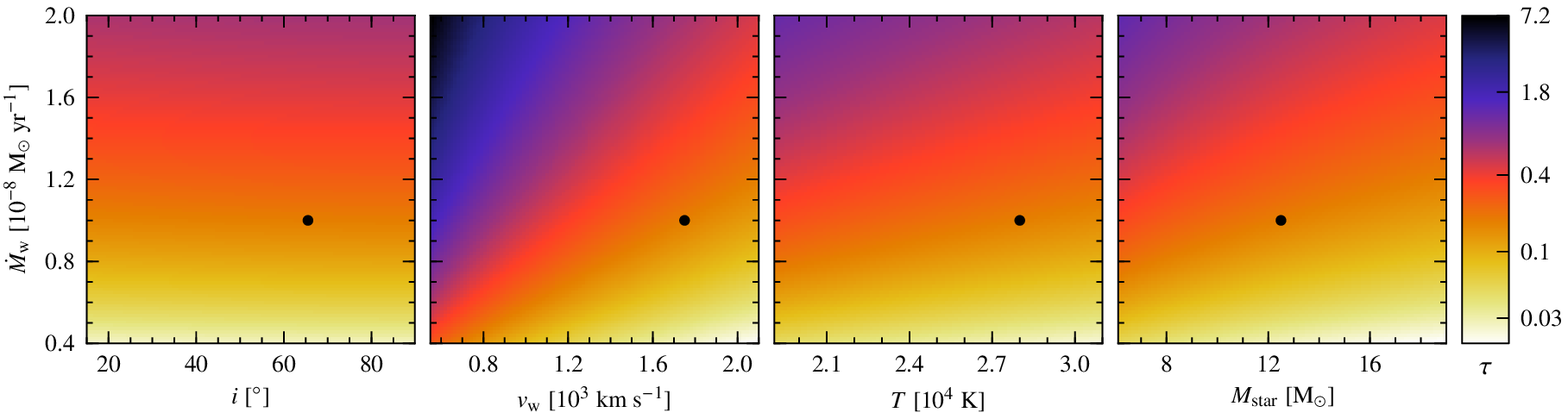}
\caption{Optical depth in logarithmic scale (colour bar) for different combinations of stellar parameters at orbital phase 0.54. The dependence with the system inclination, $i$, in the wide range 15--90$\degr$ \citep{casares05a} is very low at this orbital phase, although it is strong around periastron. The terminal wind velocity is plotted between 600~km~s$^{-1}$ (wind velocity at the equator in \cite{mcswain11}) and 2\,000~km~s$^{-1}$. The expected stellar temperature ranges from 21\,500~K \cite{mcswain11} to 30\,000~K. \cite{casares05a} estimated the stellar mass to be between 10 and 15~M$_{\odot}$. The black circles mark the values adopted in this paper.}
\label{fig:tauranges}
\end{figure*}

With these opacities we have estimated the absorbed pulsed flux densities as a function of the unabsorbed pulsed flux density $S_0$ at 1.3\,GHz,
%-------------------------------------------------------------------
\begin{equation}
S_\nu=S_0\rm{e}^{-\tau}\ ,
\label{eq:abs}
\end{equation}
%-------------------------------------------------------------------
at different frequencies and for several values of the spectral index ($\alpha$, defined as $S$\,$\propto$\,$\nu^\alpha$): $\alpha$\,$=$\,$-0.5$ (a very flat spectrum), $-1.0$ (an intermediate value) and $-2.0$ (a steep spectrum). In Fig.~\ref{fig:abs} we show the expected pulsed flux densities for $\alpha$\,$=$\,$-1.0$ for different frequencies. The modelled flux density variability of the pulsed emission at a given frequency is entirely due to the change in the the optical depth along the orbit. We conclude that the best orbital phase range to try to detect pulsations in \lsi\ is $\sim$\,0.6$-$0.7, and the highest absorbed flux densities assuming $\alpha$\,$=$\,$-1.0$ would be received at $\sim$\,0.5--2\,GHz. In Fig.~\ref{fig:tauranges} we plot the opacity for typical values of the stellar parameters. We can see that, under the conditions considered here, the opacity is around or below 1 at the orbital phase range mentioned above.

%-------------------------------------------------------------------
%-------------------------------------------------------------------
%-------------------------------------------------------------------

\section{Observations} \label{sec:observations}

The Giant Metrewave Radio Telescope (GMRT) is an array of 30 fully steerable 45\,m diameter parabolic dish antennas. It is located at 19$\degr$6$\arcmin$\,N,~74$\degr$3$\arcmin$\,E, at a height of 650\,m above the sea level, around the village of Khodad, 80\,km north of Pune (India). Fourteen antennas are distributed in a 1000\,m central core and the remaining are arranged along three $\sim$\,14\,km-long arms.

Phased array observations were performed around midnight between 2009 July 2 and 3 (UTC), centred at phase 0.54. Three observations of \lsi\ were carried out, lasting 171 minutes (21:31:45--23:22:16~UT), 123 minutes (00:52:04--03:01:02~UT) and 34 minutes (03:09:35--03:43:06~UT), i.e., a total of 5.47 hours. The array was phased on 3C48 twice: the first time before any pulsar observation, and the second time between the first and the second pointing on \lsi . We decided to observe at 1280\,MHz to minimise the effect of absorption (see Sect.~\ref{sec:absorption}) and avoid the pulse broadening (see Appendix~\ref{app}) and radio frequency interferences (RFI), which are more often present at lower frequencies. Because we aimed for a good phase coherence between antennas, we did not use the farthest three antennas in each arm. Data from two polarisations were taken in a bandwidth of 32\,MHz centred at 1280\,MHz divided into two subbands, the upper side band (USB, from 1280 to 1296\,MHz) and the lower side band (LSB, 1264 to 1280\,MHz), each split into 256 channels. Three back-ends were used simultaneously: the multi-bit phased array back-end (PA), which is specifically conceived for pulsar observations, at USB, the polarimeter (PMT) at USB and the PMT at LSB. This redundancy is convenient because each back-end introduces different noise features and this allows cross-checking. The PA sampling time was chosen to be $t_\mathrm{samp}$\,$=$\,$128\,\mathrm{\mu s}$, and $t_\mathrm{samp}$\,$=$\,$256\,\mathrm{\mu s}$ for both PMTs.

With the aim of checking the performance behaviour and data quality, we observed four known pulsars: \object{PSR~B0329+54}, \object{PSR~B1937+21}, \object{PSR~B2022+50}, and \object{PSR~B2035+36}. This group of pulsars covers a wide range of the parameters space, e.g. periods ranging from 1.56 to 715\,ms and fluxes from 0.8 to 203\,mJy. In Table~\ref{tab:ATNF}, obtained from the ATNF Pulsar Catalogue\footnote{http://www.atnf.csiro.au/research/pulsar/psrcat} \citep{manchester05}, we show the period ($P$), the dispersion measure ($DM$), the pulse width at 50\% ($W_{50}$) and 10\% ($W_{10}$) of the flux density peak, the mean flux density at 1400\,MHz ($S^\mathrm{mean}_{1400}$) and the spectral index.

%----------------------------------------
\begin{table}[t!] %Table 1
\centering
\caption{Parameters of the four known pulsars from ATNF when available. $P$ is the period, $DM$ is the dispersion measure, $W_{50}$ and $W_{10}$ are the width of pulse at 50\% and 10\% of the peak, respectively, $S^\mathrm{mean}_{1400}$ is the mean flux density at 1400\,MHz and $\alpha$ is the spectral index defined as $S$\,$\propto$\,$\nu^\alpha$.}
\label{tab:ATNF}
\begin{tabular}{ccccccc}
\hline
\hline
Pulsar Name 	& $P$   &  $DM$					& $W_{50}$ 	& $W_{10}$	& $S^\mathrm{mean}_{1400}$ 	& $\alpha$ \\
				  	& (ms)	& ($\mathrm{pc\,cm^{-3}}$)& (ms) 	   		& (ms) 	 		& (mJy)  			&         \\
\hline
B0329+54    		& 714.5	& 26.833      			& 6.6     		&  31.4   		&   203  			&  $-$1.6 \\
B1937+21    		& 1.558 & 71.0398     			& 0.063    		&  0.19   		&    10  			&   ... \\
B2022+50   			& 372.6	& 33.021      			& 4.7      		&  14.9   		&   2.2  			&  $-$0.8 \\
B2035+36    		& 618.7	& 93.56	    			& 10.7     		&   ...	 		&   0.8  			&  $-$1.6 \\
\hline 
\end{tabular}
\end{table}
%----------------------------------------
%----------------------------------------
\begin{table} % Table 2
\centering
\caption{Expected (computed with Eq.~(\ref{eq:SNR})) over obtained signal-to-noise ratios (S/N) for the four known pulsars.}
\label{tab:obsSNR}
\begin{tabular}{cccc}
\hline
\hline
pulsar name & $(S/N)_\mathrm{expected}$ & $(S/N)_\mathrm{obtained}$ & $\frac{(S/N)_\mathrm{expected}}{(S/N)_\mathrm{obtained}}$ \\
\hline
&&&\\
 \multicolumn{3}{l}{\bf PA USB}&\\
B0329+54     &  9241.2  &  179.1	  &	51.60      	\\
B1937+21     &  169.81  &  31.3 	  &	5.43      	\\
B2022+50     &  64.737  &  28.9 	  &	2.24      	\\
B2035+36     &  32.219  &  18.3 	  &	1.76      	\\
&&&\\
 \multicolumn{3}{l}{\bf PMT USB}&\\
B0329+54     &  9618.6  &  191.1	  &	50.33  		\\
B1937+21     &  158.33  &  49.8 	  &	 3.48  		\\
B2022+50     &  64.737  &  28.8		  &	 2.25  		\\
B2035+36     &  32.884  &  18.5 	  &	 1.78  		\\
&&&\\
 \multicolumn{3}{l}{\bf PMT LSB}&\\
B0329+54     &  8606.1  &  167.5	  &	51.38  		\\
B1937+21     &  141.27  &  64.3	  	  &	 2.40  		\\
B2022+50     &  59.499  &  11.8	  	  &	 5.04  		\\
B2035+36     &  30.181  &  12.3		  &  2.45   	\\
&&&\\
\hline 
\end{tabular}
\end{table}
%----------------------------------------

%Signal-to-noise ratios
We used the four known pulsars to compare the expected and the obtained $S/N$ in a manner similar to that described in \citet{joshi09}. In Table~\ref{tab:obsSNR} we show the expected and obtained $S/N$ in our observations, as well as the ratios between both values, for each of the four known pulsars. The expected $S/N$ were computed using
%----------------------------------------
\begin{equation}
S/N= \displaystyle\frac{S^\mathrm{mean}}{\frac{\beta T_\mathrm{sys}}{GN_\mathrm{a}\sqrt{N_\mathrm{p} t\Delta\nu}}\sqrt{\frac{D}{1-D}}}\ ,
\label{eq:SNR}
\end{equation}
%----------------------------------------
where $S^\mathrm{mean}$ is the mean flux density in Jy, $\beta$ is a dimensionless constant that accounts for the losses in the digitisation of the received analogue signal, $T_\mathrm{sys}$\,$=$\,$T_\mathrm{receiver}+T_\mathrm{sky}+T_\mathrm{ground}$ is the total system temperature in Kelvin, $G$ is the single antenna gain in $\mathrm{K}\,\mathrm{Jy}^{-1}$, $N_\mathrm{a}$ is the number of antennas, $N_\mathrm{p}$ is the number of orthogonal polarisations, $t$ is the integration time in s, $\Delta\nu$ is the bandwidth in Hz, and $D$ is the duty cycle defined as $D$\,$=$\,$\frac{W}{P}$, where $W$ is the width of the pulses and $P$ their periodicity.

As explained above, we used $N_\mathrm{p}$\,$=$\,2 and $\Delta\nu$\,$=$\,16\,MHz. We considered $\beta$\,$=$\,$1/0.8$, $T_\mathrm{sys}$\,$=$\,73\,K and $G$\,$=$\,$0.22$\,$\mathrm{K}\,\mathrm{Jy}^{-1}$ following the GMRT System Parameters and Current Status\footnote{http://www.ncra.tifr.res.in/~gtac/GMRT-specs.pdf} document. The mean flux density at 1280\,MHz was computed using $S^\mathrm{mean}_{1400}$ and $\alpha$ ($S^\mathrm{mean}_{1280}$\,$=$\,$S^\mathrm{mean}_{1400}\times\left(\frac{1280}{1400}\right)^\alpha$) from Table~\ref{tab:ATNF} when $\alpha$ was available and assuming $\alpha$\,$=$\,$-1.8$ for \object{PSR~B1937+21}. We used 20 antennas at USB and 18 at LSB during the observations before 3:02~UT (\object{PSR~B1937+21}, \object{PSR~B2022+50}, \object{PSR~B2035+36} and the first two observations on \lsi) and one antenna less for the remaining observations (\object{PSR~B0329+54} and the last observation on \lsi). The duty cycle was computed as $D$\,$=$\,$\frac{W_{10}}{P}$ when $W_{10}$ was available. For \object{PSR~B2035+36} we used $D$\,$=$\,$\frac{W_{50}}{P}$ instead.

The ratio of expected $S/N$ versus obtained $S/N$ should be greater than unity because Eq.~(\ref{eq:SNR}) does not model all error sources. This ratio can be much higher than 1 if the data are affected by corruption due to RFI and deviation from phasing over time. In order to take this into account we scaled the sensitivity of \lsi\ observations using a parameter that averages the expected over obtained $S/N$ ratios, which we will call the correction factor $K$. Hence, the minimum detectable mean pulsed flux density is given by

%----------------------------------------
\begin{equation}
S^\mathrm{mean}_\mathrm{min}=K\times{\left(S/N\right)}_\mathrm{min}\times \displaystyle\frac{\beta T_\mathrm{sys}}{GN_\mathrm{a}\sqrt{N_\mathrm{p} t\Delta\nu}} \times \sqrt{\frac{D}{1-D}}\ .
\label{eq:Smin}
\end{equation}
%----------------------------------------

It can be seen in Table~\ref{tab:obsSNR} that the LSB data have a lower $S/N$ (both expected and obtained) than the USB data. This is mainly because two antennas less were used in LSB than in USB. Added to this, the PA data have a shorter sampling time than PMT (128\,$\mu$s versus 256\,$\mu$s), we conclude that the PA data are the most suitable to detect pulses from \lsi .

Table~\ref{tab:obsSNR} shows that \object{PSR~B0329+54} has a much higher value of the expected over obtained ratios compared to the other pulsars. Since \object{PSR~B0329+54} is extremely bright, this result is probably reflecting a dynamic range problem. We do not expect \lsi\ to emit extremely bright pulses, therefore we did not consider \object{PSR~B0329+54} when computing the correction factor $K$.

For pulsar \object{PSR~B2035+36} the expected over obtained ratio is the lowest in each back-end. $W_{10}$ is unknown for this pulsar; this can lead to a wrong estimation of $S/N$. Moreover, \object{PSR~B2035+36} is very faint, near detection threshold, which can yield a wrong value of the obtained $S/N$. We decided not to include \object{PSR~B2035+36} in the estimation of the correction factor $K$.

For these reasons we computed $K$ by averaging only over \object{PSR~B1937+21} and \object{PSR~B2022+50} (see values in Table~\ref{tab:obsSNR}). The result is $K$\,$\simeq$\,3.8 for the PA data and $K$\,$\simeq$\,2.9 for the PMT USB data.

\section{Analysis} \label{sec:analysis}

The data of \lsi\ were analysed using the publicly available pulsar analysis package SIGPROC \footnote{http://sigproc.sourceforge.net} adapted for the GMRT data format. The data were first examined for radio frequency interference (RFI) by estimating the distribution of powers over the whole time series. Any outliers to the expected Gaussian distribution were clipped. The resulting 256-channel data for a sideband were dedispersed to 256 trial dispersion measures (DM-integrated electron column density in the line of sight) ranging from 0 to 1000\,pc\,cm$^{-3}$. The Fourier transform of the dedispersed time series for each trial DM was then computed and up to 32 harmonics in the resulting spectrum were summed. Any periodicity above eight times the signal-to-noise ratio in the summed spectrum was recorded. After ignoring the known interference periodicities, the data were folded at candidate periodicities to produce eight 2-MHz subbands and eight subintegrations over the entire observations as well as the total average profile, which were plotted as a composite plot. These plots were manually examined for all candidate periodicities, which were all rejected because their DM profiles suggested a terrestrial origin and they were not visible in all subintegrations (i.e., they were not present during all observations). Hence, no pulsed emission from \lsi\ was detected, which places a strict upper limit on the absorbed pulsed emission from this binary system.

\section{Results and implications} \label{sec:results}

\begin{figure}
\includegraphics
[width=\columnwidth]
{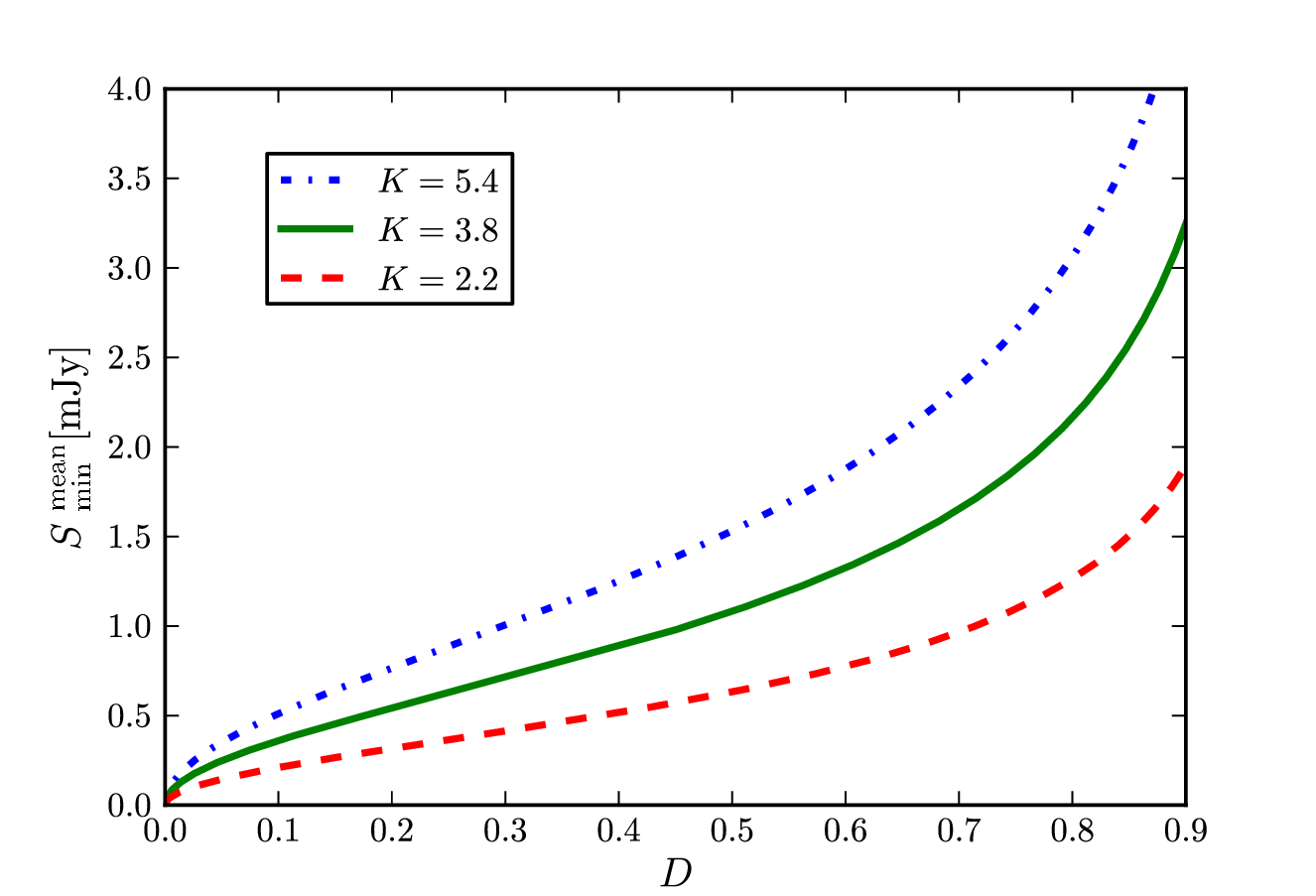}
\caption{Flux density upper limits in mJy in the direction of \lsi\ obtained with Eq.~(\ref{eq:Smin}) for increasing values of the duty cycle and for different values of the correction factor $K$ for the PA data.}
\label{fig:ul}
\end{figure}

A pulsed signal with expected DM for \lsi\ was not detected. Considering $K$\,$=$\,3.8, $t$\,$=$\,3\,hours, a ${\left(S/N\right)}_\mathrm{min}$\,$=$\,8, and a canonical duty cycle of 10\% in Eq.~(\ref{eq:Smin}) we obtained an upper limit of 0.38\,mJy for the radio pulse mean flux density from a putative pulsar with $P$\,$>$2\,ms in the direction of \lsi\ for the PA USB data from GMRT. In Fig.~\ref{fig:ul} we show the minimum mean flux density detectable in our observations for different values of the duty cycle and the correction factor $K$. For the PMT USB data the correction factor is $K$\,$\simeq$\,2.9 and the corresponding upper limit is 0.27\,mJy for pulses with $P$\,$>$5\,ms.

%----------------------------------------
%----------------------------------------

We show in Fig.~4 our obtained upper limit of 0.38\,mJy at 1.28\,GHz and the corresponding upper limits as a function of frequency for different spectral indexes including absorption (solid lines) and without absorption (dashed lines). In this plot we also show the upper limits obtained by \cite{mcswain11}, even though the observations were not performed at the same orbital phase and the sensitivities were computed using different methods. We plotted potential ten-hour sensitivities for five observatories for future observations planning: the Low Frequency Array (LOFAR) considering the HBA; the Green Banck Telescope (GBT) using the GBT Ultimate Pulsar Processing Instrument (GUPPI) and, for comparison, the GBT350 Survey sensitivity; the Effelsberg 100-m Radio Telescope considering the surveys presented in \citet{klein04}; the Expanded Very Large Array (EVLA); and the Atacama Large Millimeter/submillimeter Array (ALMA). For EVLA and ALMA, we considered a putative pulsar back-end with a full bandwidth and a reduced bandwidth ($\Delta\nu$\,$=$\,800\,MHz as for GUPPI), $N_\mathrm{p}$\,$=$\,2 and $\beta$\,$=$\,1. We have considered the most favourable conditions, and hence these estimates must be considered lower limits to the minimum mean flux density of a detectable pulsar in a ten-hour observation in the \lsi\ direction. The duty cycle is assumed to be $D$\,$=$\,10\% and in the case of the GBT it has been corrected of pulse broadening as explained in Appendix~\ref{app}.

%-------------------------------------------------------------------
\begin{figure*}[htb]
\begin{center}
\includegraphics[scale=1.0]{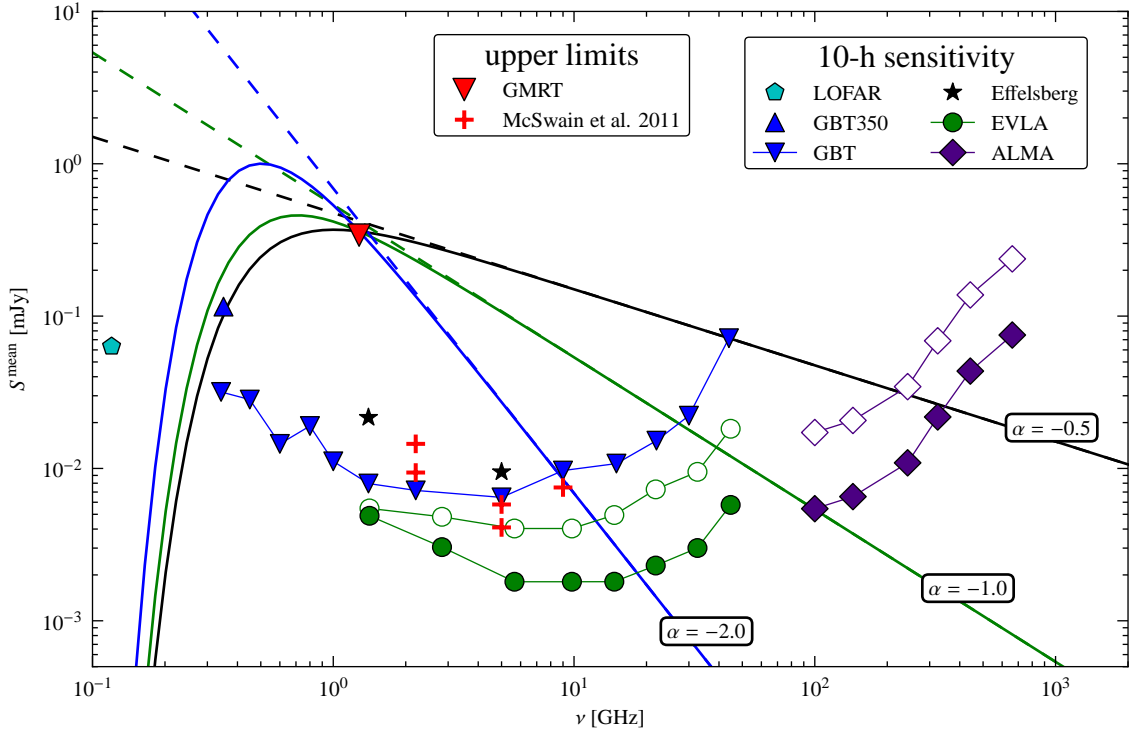}
\caption{Upper limit of 0.38 mJy extended to the radio wavelengths considering $\alpha$\,$=$\,$-0.5$ (black lines), $\alpha$\,$=$\,$-1.0$ (green lines) and $\alpha$\,$=$\,$-2.0$ (blue lines). Dashed lines do not consider any absorption, while solid lines are affected by free-free absorption using Eqs.~(\ref{eq:tau}) and~(\ref{eq:abs}) and considering the orbital position at which the absorption is lowest (phase~0.61). Note that no intrinsic low-frequency cutoff has been introduced. Ten-hour potential sensitivities are plotted for the LOFAR (pentagon), GBT (triangles), Effelsberg (stars), EVLA (circles) and ALMA (diamonds) observatories. For the two latter, empty symbols are plotted using a reduced bandwidth of $\Delta\nu$\,$=$\,800\,MHz.}
\end{center}
\label{fig:spectrum}
\end{figure*}
%-------------------------------------------------------------------

Looking at Fig.~4 it becomes evident that LOFAR operates in a zone where pulses from \lsi\ would be totally absorbed, and pulse broadening is very important if the putative pulsar is fast. In the other extreme of the radio spectrum, a ten-hour observation with ALMA would be interesting only in the case of a very flat intrinsic spectrum. It would be better to observe at the favoured frequency zone from 0.5 to 5\,GHz, especially in case of a steep spectrum. The Effelsberg telescope operates in this frequency range and would provide good sensitivities. According to Fig.~4, the EVLA would be capable to detect the lowest flux densities, but these sensitivities are computed using many assumptions, considering future planned capabilities and an unplanned pulsar back-end with assumed specifications. On the other hand, GBT enjoys the good performance of GUPPI. Furthermore, the minimum of the GBT curve is at lower frequencies, precisely at the few GHz favoured zone. In this context, we note that a high-frequency search for pulsations from \lsi\ has been performed with GBT around orbital phase 0.9, where absorption effects are not important at these frequencies according to our model, placing deep upper limits between 4.1 and 14.5 $\mu$Jy at bands S, C and X \citep{mcswain11}.

\section{Discussion and conclusions} \label{sec:discussion}
Detection of pulses is the only definitive way to confirm the pulsar nature of the compact object in \lsi . We have performed phased array observations with the GMRT at 1280\,MHz in the \lsi\ direction centred at phase 0.54, from 21:31:45 2009 July 2 UT to 03:43:06 2009 July 3 UT. No pulses have been found in the data set, with a minimum detectable flux density of around 0.38\,mJy for periods greater than 2\,ms and a duty cycle of 10\%. If \lsi\ had the same duty cycle as \psr\ (63\%), the upper limit would be 1.3\,mJy (see Fig.~\ref{fig:ul}), which is below the $\sim$\,5\,mJy mean flux density of this pulsar at this frequency \citep{johnston05}. These results give a guide for future observations planning, which would be best at $\sim$\,0.5--5\,GHz because the upper limit at 1.28\,GHz implies deep upper limits at high frequencies in case of a steep spectrum, as shown in Fig.~4. Following the discussion from Sect.~\ref{sec:absorption}, the orbital phases at which the flux is higher are $\sim$\,0.6$-$0.7, and at these phases the higher flux densities are received at $\sim$\,0.5--2\,GHz assuming a spectral index $\alpha$\,$=$\,$-1.0$. In addition to this, below 0.5\,GHz the pulse broadening due to the ISM scattering is very important for fast pulsars (see Appendix~\ref{app} for more details). Therefore, long ($\sim$\,ten hours) observations at $\sim$\,0.5--5\,GHz with Effelsberg, EVLA or more likely GBT directly pointing to \lsi\ around phases 0.6$-$0.7 are recommended to significantly improve these upper limits or have a good chance to detect the pulses. \citet{mcswain11} performed a pulsar search in \lsi\ using GBT at a similar frequency range (bands C, S and X) and a different orbital phase (0.9), placing deep upper limits (4.1--14.5\,$\mu$Jy).

In this argumentation we are not considering the possibility of the pulsar beam not pointing close to our line of sight. This issue is less problematic at X-rays, because the beam is usually wider than the radio one. However, no X-ray pulsations have been detected so far from \lsi . A deep search performed with \textit{Chandra} observations provided $3\sigma$ upper limits on the pulsed fraction of the total X-ray emission ranging between 7 and 15\% in the 0.3$-$10 keV energy band \citep{rea10}. It should be mentioned that the TeV binary system hosting \psr\ does not show X-ray pulsations during its periastron passage \citep{chernyakova06}, with an upper limit of 15\% on the pulsed fraction \citep{rea10}, yet it is known to be a pulsar beamed at Earth.

An alternative way to experimentally prove the presence of a pulsar in \lsi\ is finding HE pulsations. As mentioned in Sect.~\ref{sec:intro}, the HE emission detected by \textit{Fermi}/LAT could be explained by inverse Compton scattering of photon fields in a striped pulsar wind model, which predicts pulsed and variable HE emission \citep{petri11}.

A better orbital solution would improve the opacity estimation and the knowledge of the acceleration suffered by the putative pulsar and therefore the chances of resolving the pulses in observations along a large part of the orbit. In any case, because of the opacity of the binary system and the possibility of the putative pulsar not beaming towards Earth, it might be impossible to detect pulses even with infinite sensitivity.

\begin{acknowledgements}

We thank the staff of the GMRT who have made these observations possible. GMRT is run by the National Centre for Radio Astrophysiscs (NCRA) of the Tata Institute of Fundamental Research (TIFR). A.C., J.M.P., J.M., V.Z. and M.R. acknowledge support by the Spanish Ministerio de Ciencia e Innovaci\'on (MICINN) under grants AYA2010-21782-C03-01 and FPA2010-22056-C06-02. A.C. acknowledges support by Comissionat per a Universitats i Recerca of the DIUE of Generalitat de Catalunya and European Social Funds through grant 2010FI\_B 00291. J.M.P. acknowledges financial support from ICREA Academia. J.Mold\'on acknowledges support by MICINN under grant BES-2008-004564. V.Z. was supported by the Spanish MEC through FPU grant AP2006-00077. J.Mart\'{\i} acknowledges support by grant AYA2010-21782-C03-03 from the MICINN, research group FQM-322 and excellence grant FQM-5418 from the Consejer\'{\i}a de Innovaci\'on, Ciencia y Empresa (CICE) of Junta de Andaluc\'{\i}a and FEDER funds. M.R. acknowledges financial support from MICINN and European Social Funds through a Ram\'on y Cajal fellowship.

\end{acknowledgements}

\bibliographystyle{aa}
\bibliography{art}

\appendix
\section{Pulse width and broadening}
\label{app}

The ISM electron density variations cause a multipath scattering reflected in a temporal broadening of the pulses at high DMs and low frequencies (see for instance \citealt{cordesandrickett98,cordesandlazio02,lambert99}). In our case we have corrected the duty cycle in Eqs.~(\ref{eq:SNR}) and (\ref{eq:Smin}) using the following method. We have computed the flux density as a function of time, $S(t)$, convolving an intrinsic pulse profile, $S_\mathrm{int}(t)$, and a broadening function $b(t)$. Only with this purpose we considered an intrinsic Gaussian pulse profile with standard deviation $\sigma$,
%----------------------------------------
\begin{equation}
S_\mathrm{int}(t)=\frac{PS^\mathrm{mean}}{\sqrt{2\pi\sigma^2}}\mathrm{e}^{-\frac{t^2}{2\sigma^2}}\,,
\label{eq:Sint}
\end{equation}
%----------------------------------------
with a period of $P$\,$=$\,10\,ms, an intrinsic duty cycle of $D_\mathrm{int}$\,$=$\,10\% and hence an intrinsic pulse width of $W_\mathrm{int}$\,$=$\,1\,ms. In order to link $\sigma$ and $W_\mathrm{int}$ in this context, we considered $W= W_{10}$ (the pulse width at 10\% of the peak). Therefore, $\sigma$\,$=$\,${W_{10,\,\mathrm{int}}}/{\sqrt{8\ln 10}}$.

The temporal pulse broadening is parametrised with the timescale $\tau_\mathrm{d}$, defined as the first moment of the broadening function, roughly a one-sided exponential, determined by deconvolving the intrinsic and measured pulse shapes (see for instance \citealt{cordesandlazio02,cordesandrickett98}). We effectively considered a one-sided exponential for the pulse broadening function $b(t)$:
%----------------------------------------
\begin{equation}
b(t) = \left\{
\begin{array}{ll}
\frac{1}{\tau_\mathrm{d}}\mathrm{e}^{\frac{-t}{\tau_\mathrm{d}}} & t\geq 0 \\
0 & t< 0.
\end{array}
\right.
\label{eq:b}
\end{equation}
%----------------------------------------
Hence, $\tau_\mathrm{d}$ is the width of the pulse to $1/\mathrm{e}$ of the maximum pulse amplitude if the intrinsic pulse shape is a delta function. We have obtained the values of $\tau_\mathrm{d}$ at several frequencies using the NE2001 electron density model \citep{cordesandlazio02} in the direction of \lsi\ and considering a distance to the source of $d$\,$=$\,$2.0\pm0.2$\,kpc.

Finally, the measured pulse profile $S(t)$ is the convolution of $S_\mathrm{int}(t)$ and $b(t)$:
%----------------------------------------
\begin{eqnarray}
S(t)&=&S_\mathrm{int}(t)\otimes b(t) = \nonumber \\
    &=& \frac{P S^\mathrm{mean}}{2\sigma\tau_\mathrm{d}} \times \mathrm{e}^{\frac{-t}{\tau_\mathrm{d}}+\frac{\sigma^2}{2\tau_\mathrm{d}^2}}\times \mathrm{erfc} \left( -\frac{t}{\sqrt{2}\sigma}-\frac{\sigma}{\sqrt{2}\tau_\mathrm{d}} \right)\ .
\label{eq:god}
\end{eqnarray}
%----------------------------------------

%----------------------------------------
\begin{figure}[htb]
\includegraphics[scale=1.07]{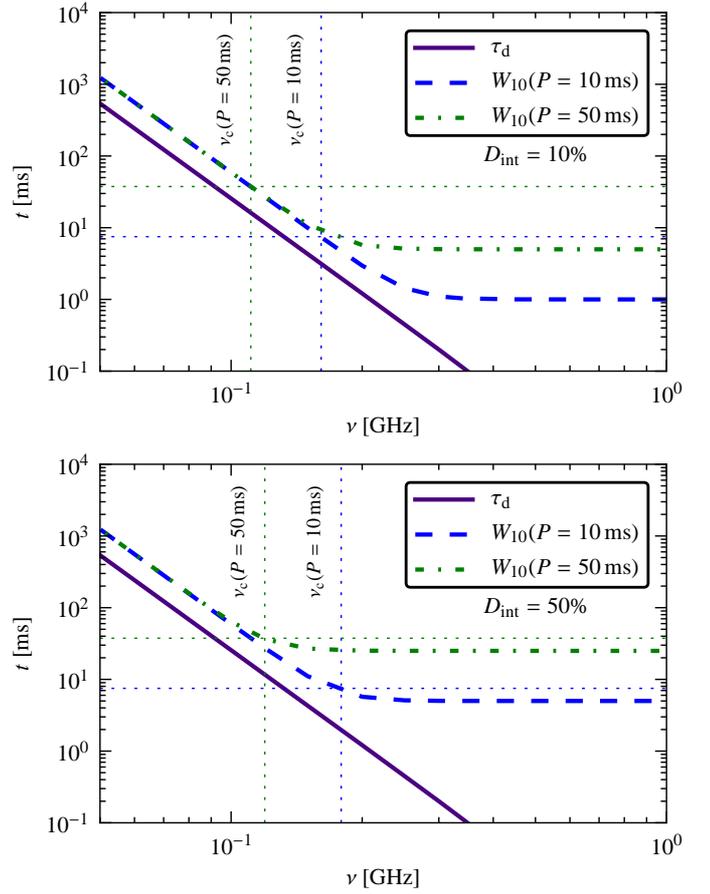}
\caption{Pulse widths broadened due to ISM scattering in the frequency range at which the role of pulse broadening turns from crucial to negligible. The resulting width $W_{10}$ is plotted considering $P$\,$=$10\,ms (blue dashed lines) and $P$\,$=$50\,ms (green dash-dotted lines) for intrinsic duty cycles $D_\mathrm{int}$\,$=$10\% (top panel) and $D_\mathrm{int}$\,$=$50\% (bottom panel). The pulse broadening timescale $\tau_\mathrm{d}$ is plotted with a solid indigo line. Horizontal lines mark 75\% of period, and vertical lines indicate the $\nu_\mathrm{c}$ where the pulse widths cross this limit.}
\label{fig:broad}
\end{figure}
%----------------------------------------

In Fig.~\ref{fig:broad} we show the resulting pulse widths with $P$\,$=$\,10\,ms and 50\,ms, and $D_\mathrm{int}$\,$=$\,10\% (top panel) and 50\% (bottom panel). Vertical lines indicate the frequencies $\nu_\mathrm{c}$ at which $D$\,$=$\,75\% and thus the pulse is hardly detectable. For an intrinsic duty cycle of 10\%, above $\sim$\,0.5\,GHz the effect is negligible. We conclude that the pulse broadening does not affect our observations at 1.28\,GHz. In Fig.~4 this correction is used considering $P$\,$=$\,10\,ms to the points below 1\,GHz. According to this result, at 120\,MHz (the High Band of LOFAR) this presumed pulsar would not be detectable because the pulse width would be greater than the period. For this reason the LOFAR point is not corrected for pulse broadening. If the compact object in \lsi\ is a fast ($P$\,$\sim$\,10\,ms) pulsar, observations above 0.5\,GHz are recommended to avoid the pulse broadening by the ISM.

\end{document}